\documentclass[sigconf, screen]{acmart}

\AtBeginDocument{%
  \providecommand\BibTeX{{%
    \normalfont B\kern-0.5em{\scshape i\kern-0.25em b}\kern-0.8em\TeX}}}

\setcopyright{acmcopyright}
\copyrightyear{2026}
\acmYear{2026}
\acmDOI{XXXXXXX.XXXXXXX}

\acmConference[EASE 2026]{The 30th International Conference on Evaluation and Assessment in Software Engineering}{9–12 June, 2026}{Glasgow, Scotland, United Kingdom}

\acmPrice{15.00}
\acmISBN{978-1-4503-XXXX-X/18/06}

\usepackage{todonotes}

\begin{document}

\title{Enhancing a gamified tool for UML modeling education}

\author{Giacomo Garaccione}
\affiliation{%
  \institution{Politecnico di Torino}
  \city{Turin}
  \country{Italy}
}
\email{giacomo.garaccione@polito.it}

\author{Riccardo Coppola}
\affiliation{%
  \institution{Politecnico di Torino}
  \city{Turin}
  \country{Italy}
}
\email{riccardo.coppola@polito.it}

\author{Luca Ardito}
\affiliation{%
  \institution{Politecnico di Torino}
  \city{Turin}
  \country{Italy}
}
\email{luca.ardito@polito.it}

\renewcommand{\shortauthors}{Garaccione et al.}

\begin{abstract}
  Unified Modeling Language (UML) Use Case and Class Diagrams are fundamental modeling notations in Software Engineering (SE) education due to their importance for requirements and model-based engineering, yet their relevance is underestimated by students, who tend to dismiss the topic as secondary. Gamification has been adopted to make modeling education more appealing, but existing tools focus almost exclusively on class diagrams, leaving support for use cases and other notations unexplored. In 2025, we designed UMLegend, a gamified tool for class diagrams that offered dynamic feedback to help students learn correct modeling practices and multiple long-term mechanics to increase engagement, and performed a study with the tool. With this paper, we describe how we enhanced UMLegend following the results of the experiment so that it can support more modeling languages, with use case diagrams being added to the type of available exercises in the tool. The revised version has been refactored to have a modular architecture, to make it easier to add other software engineering topics and additional modeling notations. We also describe the potential impact we expect the new version to have, and outline a longitudinal study we intend to perform in 2026 where we will assess whether long-term UML gamification leads to improved student performance.
\end{abstract}



\maketitle

\section{Introduction}

Unified Modeling Language (UML) is a graphical notation that is one of the most commonly used in software engineering practices, playing a pivotal role for the visual representation of multiple aspects in the software engineering life-cycle such as requirements definition and design. 

Among the many types of diagrams that it defines, use case and class diagram stand out as key topics due to their versatility in requirements engineering, where they are adopted to represent functional requirements, intended system usage, and key implementation, as well as in model-based engineering, where they are used as an artifact on which implementations depend. 

UML is commonly taught in SE curricula due to its importance, although students perceive it as a somewhat secondary topic, dedicating less attention to it compared to coding or testing activities, resulting in turn in the production of diagrams with errors and limited ability to represent the intended domain, reducing the positive impact of UML models. Educators thus need to identify techniques and approaches that can increase student engagement and commitment, and apply such techniques to UML education.

Gamification refers to the use of game design elements in non-game contexts with the aim of increasing user engagement and motivation \cite{deterding2011gamification}, and is a technique that has emerged over time as an effective means to increase student engagement and motivation in activities that are usually perceived as unappealing or uninteresting such as software design and testing, as well as requirements engineering activities \cite{alhammad2018gamification}.

Gamification has been applied to many software engineering disciplines, including UML education with mechanics such as points, badges, achievements, levels, and leaderboards, with positive results in terms of student engagement and performance \cite{Jurgelaitis2018UsingGF, jurgelaitis2019implementing}. 

We previously introduced UMLegend \cite{garaccione2026gamification}, a gamified tool for UML class diagram exercises that allowed dynamic exercise solving with feedback dependent on the students' produced diagrams, and support for long-term mechanics to allow for continuous classroom use aimed at increasing participation in practical sessions.

We conducted an empirical experiment at Politecnico di Torino with 280 students to assess its impact on student performance, demonstrating that by gamification brought an increase in diagram correctness, together with positive student perception of the experience. However, the study was performed in a single session, preventing any generalization regarding longitudinal impact; furthermore, the tool presented certain limitations such as exclusively supporting class diagrams as the modeling notation, and having a limited evaluation engine for the exercises.

As a consequence, we continued developing the tool so that it could support more classes of diagrams and reworked both the game engine and the evaluation architecture to make using the tool more pleasant, with the goal of creating a multi-purpose gamified environment for software engineering education. This paper describes the revised version of the tool, for which we have made the following changes:

\begin{itemize}
    \item Most game mechanics have been reworked following feedback from the original experiment, resulting in a less penalizing experience with clearer feedback.
    \item The evaluation engine has been extended to support more variability in the space of available solutions, and the ability for teachers to create references has been enhanced.
    \item Use case diagrams have been added as a supported modeling notation for both the gamified mechanics, the evaluation engine, and the reference solution creator.
    \item The tool architecture has been refactored in a modular fashion to allow for easy insertion of other topics, including SE ones not related to modeling, with the end goal of making the tool architecture available as an open-source platform for the educational and research community.
\end{itemize}

We intend to perform a longitudinal study throughout spring 2026 by adopting the revised UMLegend into a software engineering course, to assess whether gamification can improve student performance for both class and use case modeling exercises.

The remainder of this paper is structured as follows: Section 2 presents an overview of gamification applied to software engineering disciplines, with a focus on model-based education and requirements engineering, Section 3 outlines the changes made in the revised version of UMLegend, while Section 4 describes the expected impact of the new version of the tool and the plans for the experiment we intend to perform.


\section{Background}

In recent years, gamification has been adopted in multiple educational contexts to stimulate student participation and improve learning performance through the implementation of engaging game mechanics and elements. Multiple empirical studies have investigated the effectiveness of educational gamification, showing that it can have a positive effect on motivation and engagement \cite{hamari2014does}, despite different contexts and implementation strategies performing better than others; gamification has also shown to be beneficial for both cognitive and behavioral engagement when the mechanics' design connects them to specific learning objectives \cite{seaborn2015gamification}.

Applications focused on software engineering education showed that gamification can lead to improvements in student engagement and learning performance for coding assignments \cite{ibanez2014gamification, marin2018empirical}, team-based development\cite{bilal2014enhancing}, requirements definition \cite{cruz2025experimental}, and testing activities \cite{fraser2019gamifying, straubinger2024improving}.

Despite the widespread use of gamification in software engineering education, examples of its application to software modeling or model-based engineering activities are still quite rare, with most of them focusing on UML class diagrams \cite{ junior2021modelgame, marin2018learning, Cammaerts2023ModelDefendersAN} through various mechanics such as competition, feedback, points, rewards, and achievements. 

However, gamification focused more on use case diagrams remains a relatively unexplored field. Jurgelaitis et al. \cite{Jurgelaitis2018UsingGF} discussed the design of an Information System course where multiple-choice exercises and questions on multiple UML notations (including use case diagrams) were served through the online platform Moodle, and enhanced these exercises with experience points, badges, and leaderboards. An application of their design was positively received by students \cite{jurgelaitis2019implementing}, with around half of them managing to complete the use case portion of the course.

Other tools exist that cover multiple UML notations without an explicit focus on use cases such as Papygame \cite{bucchiarone2023gamifying}, an extension for the Papyrus modeling platform where completing exercises awards progression points and feedback for errors, and GaMoVR \cite{yigitbas2024gamovr}, a virtual-environment game where the UML components can be directly rearranged to compose a valid diagram, with levels, experience points, and unlockable rewards as the main mechanics. However, the evaluation of these tools has focused mainly on class diagrams, meaning that the application of gamification to use case diagrams remains unexplored.

Furthermore, the existing solutions for gamifying multiple UML notations present static exercises where students have to select correct options among a set of available elements, and no tool exists where students can directly attempt to draw their personal solution on a modeling canvas. The absence of such a tool was our original motivation behind the development of UMLegend, and our intent with this paper is to extend it with a dedicated module for gamifying use case diagram modeling, while maintaining the original philosophy of the tool.





\section{Tool Features}

\subsection{Game Mechanics}

Following the original implementation of UMLegend, the new version keeps the game mechanics separated into two categories: long-term mechanics that support the users' growing skill in modeling languages over time and mechanics that impact how users interact in a specific exercise, reacting to the solutions produced and their content in relation to the exercise's context. 

\subsubsection{Long-term Mechanics}

All the long-term mechanics implemented in the original version of UMLegend have been kept in this revised version, although all of them have been changed to improve the overall student experience.  

For what concerns \textbf{Levels and Experience Points} still exist as a way to measure the students' natural progression and growth in modeling skill, with experience being obtained by successfully completing exercises like in the original version. The main change made is that all parameters tied to experience points can now be directly customized by teachers on a course by course basis, instead of being hard-coded values; as a consequence, teachers can define how many levels can be obtained, the experience threshold for each level, as well as the thresholds needed to obtain experience multipliers, as well as the values associated with those multipliers.

A similar change has been made for the \textbf{Avatar Customization} mechanic: the level conditions for unlocking avatar props were originally hard-coded and can now be changed directly by teachers. Furthermore, the way avatar props are displayed so that students can customize their look has been changed to show how each prop changes the avatar's style compared to the currently selected options. 

Avatars still maintain their original purpose of being shown inside \textbf{Leaderboards}, which remain the way UMLegend implements competition between students. In addition to the original rankings (by level and experience points, and by correctness score per each exercise), a third leaderboard has been added, ranking users by the number of completed exercises.

Lastly, a new long-term mechanic has been added in the form of \textbf{Completion Rewards}, that apply changes to the student's homepage after completing an exercise. The first change is that the completed exercise's corresponding block is highlighted in golden and displays the corresponding enemy's icon, giving a visual change to the exercise page and providing a satisfactory improved display; the second change consists of allowing the students to access a dedicated section of the tool
where it is possible to consult the reference solutions for the exercises that have been completed.

\subsubsection{Exercise-specific Mechanics}

Similarly to the long-term mechanics, all the original exercise-dependent mechanics have been kept in the improved version, with most of them also being changed following feedback from the original experiment; new mechanics have been added as well, but the original concept remains the same: mechanics are activated in reaction to a student's diagram being evaluated and allow for a personalized experience that guides toward learning correct modeling practices.

\begin{figure*}[h]
    \centering
    \includegraphics[width=0.70\linewidth]{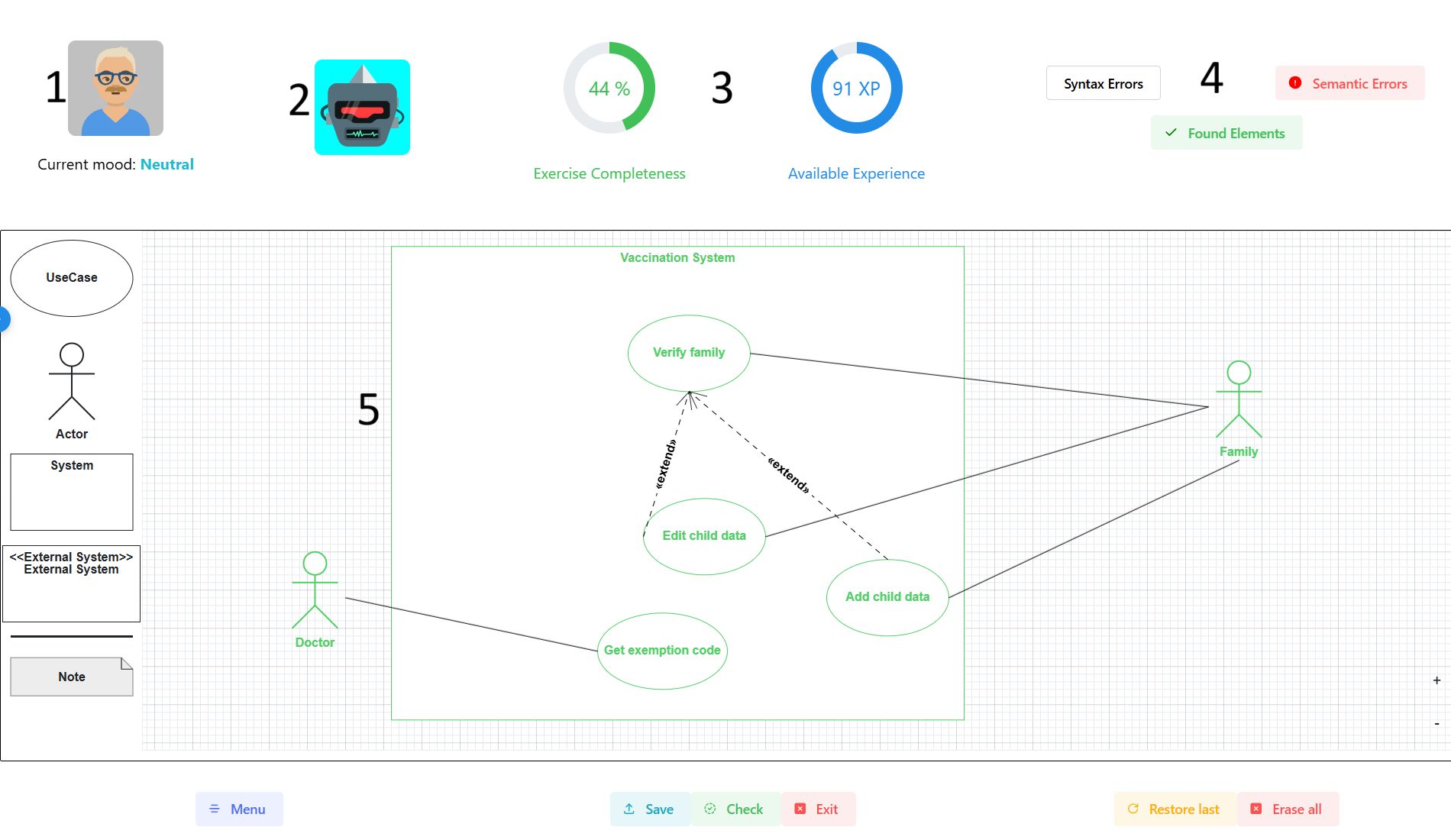}
    \caption{Exercise page showing the exercise-dependent mechanics}
    \label{fig:ex_page}
\end{figure*}

\textbf{Indicators} remain as the guiding mechanic that helps students track how well they are performing in an exercise by knowing their current completeness score and obtainable experience points. Completeness is still computed in the same way as in the original version, that is, as the count of elements in the reference solution that are considered a valid match over the total number of reference elements. The experience reward has been changed in some ways, following feedback gathered after the original experiment: one of the main complaints was that the reduction in experience felt too strict and frustrating, and we decided to address this by making it so that only new errors cause a reduction of the obtainable experience; repeating an error does not reduce the experience points anymore, and we assume that this should mitigate the frustration found in the original version. Additionally, the experience calculation still ensures that errors never reduce the obtainable value below 35\% of the total value, and correcting errors increases the experience score by the same amount it was reduced originally. Additionally, indicators have received a visual change and became progress circles, to improve their visibility (see Figure \ref{fig:ex_page}(3)).

The \textbf{Avatar Feedback} was used as a negative motivator in the original implementation, with the student's avatar becoming progressively sadder with the reductions in the obtainable experience points. The revised version has implemented two changes: first of all, the avatar's mood is updated after every check by comparing its results with those obtained in the preceding one, deciding the new mood depending on the difference between the experience values, the number of new errors, the number of fixed errors, and the difference in completeness values. With this change on the criteria that affect the avatar's mood, it is possible to have a change that makes it sadder, does not change its mood, or makes it happier, leading to three positive states, three negative ones, and a neutral one. The avatar's mood is stated in the upper left side of the exercise page, with Figure \ref{fig:ex_page}(1) showing an example of how it looks like.

To support the students' understanding of this mechanic, a new one has been implemented to allow students to easily gauge the changes between two subsequent evaluations in the form of the \textbf{Evaluation Recap}: after every check, a modal window appears showing the results and listing how many existing errors have been fixed, how many new errors have been found and, consequently, the changes in experience and correctness.

Another new mechanic that relates to the students' avatars during an exercise has been added in the form of a robotic \textbf{Boss} that visually represents the challenge associated with the exercise and taunts the students with a customizable phrase, as shown in Figure \ref{fig:ex_page}(2). The boss disappears with an animation when the exercise is completed, signifying that the student has managed to win the challenge, and it appears on the exercise's icon in the homepage as a badge of success. The boss' look and its corresponding dialogue can be customized by a teacher for each exercise in a dedicated section of the tool.

The \textbf{Completion Screen} is another mechanic that has been added in this improved version: when a student completes an exercise, a modal window appears showing the obtained experience points (with the eventual bonus multipliers) and a progress bar that shows the updated student experience. If the student has reached a new level, the list of unlocked avatar props is also shown in this window.

Two mechanics that were present in the original version (\textbf{Error Lists} and \textbf{Diagram Feedback}) have been changed to provide improved guidance and help toward successful exercise completion and correct modeling understanding due to a change in how the evaluation results are managed: the original version only considered syntactic and semantic errors as information to convey to students, while the revised version considers matching elements something relevant that students should know.

As a consequence, the tool now displays \textbf{Feedback Lists} for syntactic and semantic errors, as well as a third list for the diagram elements that are considered valid matches for elements in the reference solution. Furthermore, messages have been rewritten to ensure that the affected diagram element is mentioned explicitly, and the reasoning behind the error in relation to the reference solution is clear. Feedback lists appear in the upper right side of the exercise page, as shown in Figure \ref{fig:ex_page}(4).

\textbf{Diagram Feedback} has been changed as well, since now correct elements can also be colored in green. Coloring is done in a way that ensures correct elements can always be identified by prioritizing the green background, while errors have a red/orange text depending on whether they are semantic or syntactic ones. Figure \ref{fig:ex_page}(5) shows a diagram where this mechanic is applied and all elements are considered correct.

The revised version also keeps the same in-exercise leaderboard as in the original version, with no additional change made to said mechanic.

\subsection{Use Case Diagram Evaluation}

The main improvement of the revised version of UMLegend compared to the one used in the original experiment is the support for use case diagrams in its automated evaluator, extending the range of software modeling topics that it can cover and teach. The evaluation module for UCDs identifies the following three main elements for its intended reference solution structure: actors, use cases, and systems, distinguishing the latter between "owning" systems that are allowed to contain use cases, and "external" systems that cannot contain use cases and are intended to be used as supporting actors for the use cases to represent. 

In addition, relationships between elements are also defined in the reference solution structure, with three different relationship types: I) associations between use cases and actors/systems, used to state which actors perform which use cases or, if the corresponding flag is set, if the actor is a supporting one for the use case; II) associations between actors, representing inheritance between two actors; III) associations between two use cases, with a flag being used to state whether the association is an \textit{Inclusion} or an \textit{Extension}.

The automated evaluator works by receiving the Apollon source code of a student's diagram and parsing it so that it is converted to a data structure of the same type as the reference solution, ensuring compatibility for the comparison between the two.

The evaluation procedure's first step consists of comparing each of the reference elements (actors, use cases, system) with all diagram elements of the same type until one element with a similar enough name (computed with string similarity using Levenshtein distance and requiring at least 75\% similarity) is found, signifying a match between the two elements; following this, the evaluator iterates through the reference associations to check if, for each of them, both involved elements have a matching element in the diagram and, if that is true and there is an association between the pair, considers the relationship matched.

Once all matches have been found, the evaluator computes completeness metrics for each element type, as well as the overall completeness, and then performs two additional evaluation steps to identify syntactic and semantic errors. Syntax errors do not depend on the student's solution, since they focus on structural issues such as elements with missing or duplicate names, wrong or not allowed associations, misplaced elements such as actors inside systems or use cases outside systems.

Semantic violations, instead, depend on the result of the comparison and highlight errors such as missing matches for reference elements, matching use case associations with wrong type (e.g., an \textit{Include} in place of an \textit{Extend}) or with the wrong order of use cases, use cases belonging to the wrong system, elements with names that are not allowed, and relationships between elements that should not be included.

\begin{figure*}[t]
    \centering
    \includegraphics[width=0.75\linewidth]{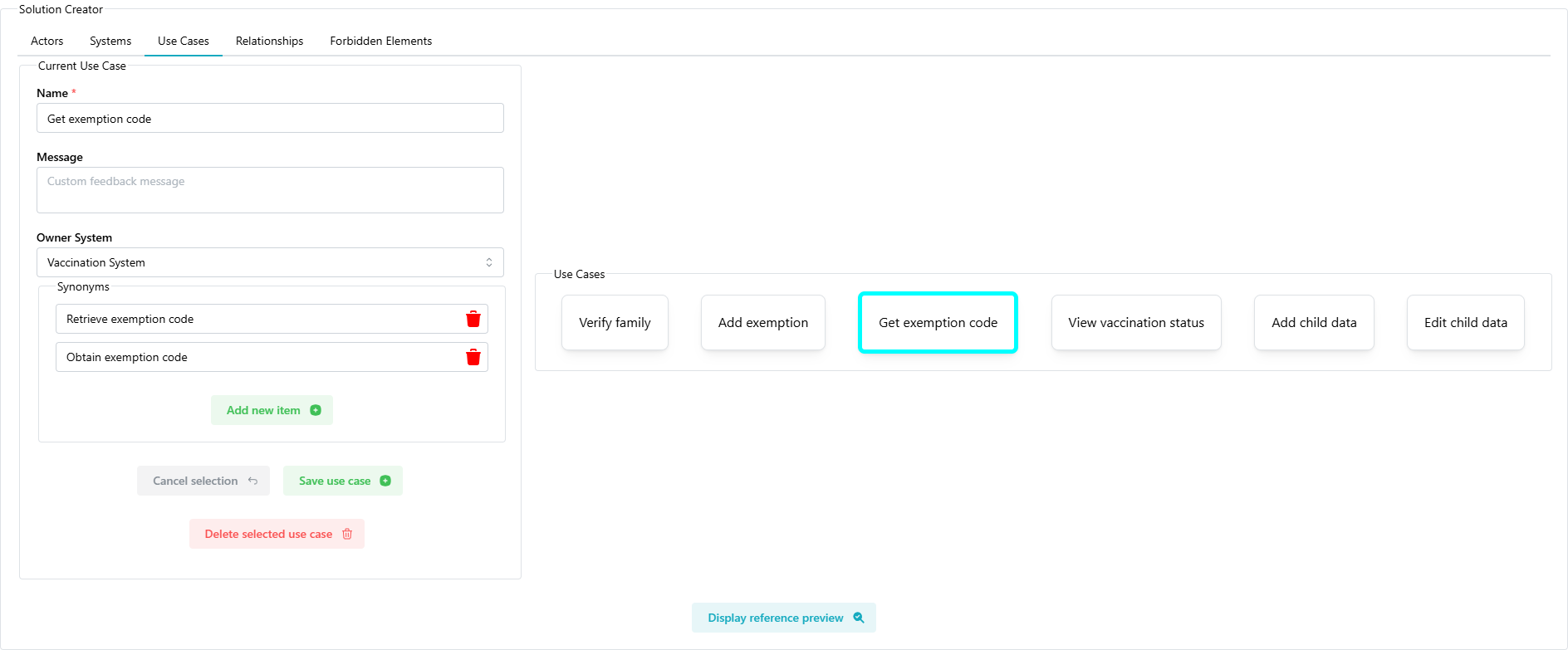}
    \caption{Form for creating a reference solution for a use case diagram, showing an editable use case}
    \label{fig:sol_form}
\end{figure*}

The reference solution structure for an exercise can be defined by teachers in two ways: drawing it by using Apollon's modeling interface, or through a form where it is possible to specify all expected elements and their details. The form allows teachers to define all elements separately and establish their relationships after creation: Figure \ref{fig:sol_form} shows how the solution form looks like when editing a use case by specifying its owning system and possible alternative names for it.

An important change compared to the original version is the support for multiple solutions to the same exercise, letting teachers cover more variability in the possible representations of a problem. Consequently, the evaluation procedure described above is performed after every check for all solutions created for an exercise, and the feedback is displayed for the solution that has achieved the highest completeness score.

\section{Planned Use and Impact}

The original experiment we conducted using UMLegend showed that gamifying UML modeling assignments results in increased diagram correctness and fewer errors, and that it is a well-received experience by students. However, the experiment was conducted as a single session, meaning that we have no way to generalize the findings to longitudinal use of the tool. Furthermore, the original version covered exclusively class diagrams, leaving out other modeling notations such as use case or deployment diagrams that are commonly used in requirements engineering and are a relevant topic for software engineering education.

The revised version has been developed to provide an educational environment that can support multiple modeling notations in a modular structure, meaning that additional notations can be included in the tool by defining the related evaluation logic and the corresponding solution creation form. As a consequence, we envision the possibility to define a basic gamified educational environment that can be extended to support different software engineering topics, not strictly related to modeling languages, that teachers can extend and deploy.

We intend to deploy the revised UMLegend and adopt it during the 2026 edition of the Software Engineering course at our university, where it will be made available for students to use for the duration of the course to solve multiple class diagram and use case diagram assignments. Our goal is to assess whether continuous use of a gamified modeling platform can lead to improved student performance by observing how student performance changes over time, as well as how student perform in modeling assignments during the course's final exam.

The results of this examination will serve as an example of how the long-term application of gamification impacts software engineering education, providing insights for the research community; additionally, we plan to make the tool available as an open-source project so that other researchers can apply it in their courses, and extend it so that it supports other topics.

\bibliographystyle{ACM-Reference-Format}
\bibliography{sample-base}

\end{document}